\documentclass[structabstract]{aa}
\usepackage{graphicx, xspace, lscape}
\usepackage{float}
\usepackage{longtable}
\usepackage{savesym}
\usepackage{amsmath}
\usepackage{txfonts}

\begin{document}
\title{A VLT/FLAMES survey for massive binaries in Westerlund~1: VI. Properties of X-ray bright  massive cluster members\thanks{This work is based on observations collected at the European Southern Observatory, 
     Paranal Observatory under programme IDs ESO 081.D-0324 and 091.D-0179}}
\author{J.~S.~Clark\inst{1} \and B.~W.~Ritchie\inst{1,2}  
      \and I.~Negueruela\inst{3}}
\institute{
$^1$School of physical sciences, The Open University, Walton Hall, Milton Keynes MK7 6AA, United Kingdom \\
$^2$Lockheed Martin Integrated Systems, Building 1500, Langstone, Hampshire, PO9 1SA, UK \\
$^3$Departamento de F\'{\i}sica Aplicada, Facultad de
Ciencias, Universidad de Alicante, Carretera San Vicente del Raspeig s/n,
E03690, San Vicente del Raspeig, Spain}
   \abstract{X-ray emission from massive stars was first reported four decades ago, but the precise physics governing its formation as a function of stellar properties and binarity remains to be fully understood. With the recent suggestion that such objects may be important sites of cosmic ray production, a better understanding of their high-energy properties   is particularly timely.}
{The young massive cluster Westerlund 1 provides an 
ideal testbed for understanding this emission, with over 50 
cluster members detected in historical X-ray observations. In the decade since these data were obtained, significant 
new multi-epoch observations of the cluster have been made, allowing a fundamental reappraisal 
of the nature of both  X-ray bright and dark stars.}
{Optical spectroscopy permits accurate classification of cluster members, while multi-epoch observations of a subset allows identification and characterisation of the binary population.}
{A total of 45 X-ray sources within Wd1 now have precise spectral classifications. Of these, 16 have been identified as candidate or confirmed massive binaries. X-ray emission is confined to O9-B0.5 supergiants, Wolf-Rayets and a small group of highly luminous 
interacting/post-interaction OB+OB binaries. Despite their presence in large numbers, no emission is seen from earlier, less evolved O stars or later, cooler B super-/hypergiants. A total of 22 stars have X-ray properties that are suggestive of a contribution from emission originating in a wind collision zone.}
{We suppose that the lack of X-ray emission from O giants is due to their comparatively low intrinsic bolometric luminosity if, as expected, they follow the canonical
$L_{\rm X}/L_{\rm bol}$ relation for hot stars. The transition away from X-ray emission for OB supergiants occurs at 
the location of the bistability jump; we speculate that below this limit, stellar wind velocities are insufficient for internal, X-ray emitting shocks to form. Our results  are consistent with recent findings that massive binaries are not uniformly brighter than single stars of comparable luminosity or spectral type, although it is noteworthy that the brightest and hardest stellar X-ray sources within Wd1 are all either confirmed or candidate massive, interacting/post-interaction binaries.

}
   \keywords{stars: evolution - supergiants - stars: Wolf Rayet - stars: binaries: general - stars: winds, outflows - X rays: stars 
   }
   \maketitle
%

\section{Introduction}

Since the advent  of X-ray astronomy hot, massive stars have been recognised as important sources of emission 
(e.g. Seward et al. \cite{seward79}), with the realisation that O stars  obeyed an empirical
X-ray/bolometric luminosity relation - $L_{\rm X} \sim 10^{-7}L_{\rm bol}$ - quickly following 
(Long \& White \cite{long},  Seward \& Chlebowski \cite{seward82}). The consensus view holds that this emission arises in hot shocked material embedded in a cooler bulk stellar wind as a results of the line driving instability inherent in radiatively driven winds (Lucy \& White \cite{lucy}, Feldmeier et al. \cite{feld}), with the temperature of shocked material consistent with the soft ($kT\sim0.6$keV) X-ray spectra of the majority of hot stars. A subset of hot stars are found to be significantly brighter and/or harder than these canonical sources; such emitters are typically found to be massive binaries, where it is thought that  additional emission arises in shocked material in a wind collision zone (WCZ; cf.  Stevens et al. \cite{stevens}, Pittard \& Dawson \cite{pittard} and refs. therein).

However the $L_{\rm X}/L_{\rm bol}$ relationship appears to break down for stars of spectral type B0-2 and later, which are found to be significantly fainter than predicted (Cassinelli et al. \cite{cassinelli}, Bergh\"{o}fer et al. \cite{bergh}). Likewise Wolf-Rayet (WR) stars also diverge from this correlation, with (moderate) emission  dependent on sub-type and binarity (e.g. Oskinova 
\cite{oskinova16}; Sect 3). Unfortunately, the physical basis for the $L_{\rm X}/L_{\rm bol}$ relationship is currently uncertain,
making such observational findings difficult to interpret. Analysis by Owocki \& Cohen (\cite{owocki}) suggests that X-ray emission driven by shocks should be sensitive to the wind density parameter (\.{M}/$v_{\rm wind}$) rather than to $L_{\rm bol}$ (although in practice the former will be somewhat sensitive to the latter). Sciortino et al. (\cite{sciortino}) show a correlation between X-ray luminosity and the total wind momentum flux, while 
Nebot G\'{o}mez-Mor\'{a}n \& Oskinova (\cite{nebot}) also find that emission scales with wind properties for main sequence and giant O stars.

Given that the X-ray flux may change the ionisation balance of the wind and hence affect mass loss determinations, an understanding of the production mechanism and its relation to underlying wind and stellar parameters is clearly required; 
especially since determination of the wind properties of massive stars is essential if one it  to constrain their evolution and feedback into their immediate environment. Moreover it has been suggested that the (interacting) winds of massive single and binary stars may serve as sources of cosmic rays, accelerating protons to PeV energies  (e.g. Cesarsky \& Montmerle \cite{CM}, Bednarek et al. \cite{bed}, Aharonian et al. \cite{aharonian}); consequently understanding the physics and prevalence of such  wind interaction zones via the X-rays they emit is important if we are to test this hypothesis.

\begin{table}
\caption{Stellar demographics of  X-ray sources within Wd1}
\begin{center}
\begin{tabular}{lcc}
\hline
\hline
Spectral    & Cluster     &  X-ray   \\
Type        & population  & detected \\
\hline
O9-9.5 III             &   27  & 1 \\
O9-9.5 II,II-III       &   12  & 1 \\
O9-9.5 Iab,Ib          &   30  & 19 \\
B0-0.5 Ia,Iab,Ib        &   23  & 6 \\
B1-1.5 Ia,Iab   & 10     & 0    \\
B2-4 Ia         & 7 & 0 \\  
B5-9 Ia$^+$     & 4 & 0 \\ 
LBV             & 1 & 0 \\
YHG+RSG         & 10 & 0 \\[2.5mm] 
O4-8 Ia$^+$     & 2 & 2 \\
B0-2Ia$^+$/WNVL & 4 & 2 \\
sgB[e]          & 1 & 1 \\
OB SB2          & 2 & 2 \\[2.5mm]
WN5-8           & 14 & 8 \\
WC8-9           & 8 & 3 \\
\hline
\end{tabular}
\end{center}
{Summary of X-ray detections by spectral type. The top panel reflects the expected evolutionary
pathway for single stars, the second the possible products of binary interaction and the third 
WRs - a structure chosen to reflect the discussion in Sect. 3. Note that the marginal X-ray detections
discussed in Sect. 3.4 are not included here.
The X-ray detections Wd1-13,  W1033 and Wd1-232  are included with the B0-2Ia$^+$/WNVL, 
O9-9.5 II-III and B0-0.5 Ia,Iab,Ib subsets, respectively. The OB2 SB2 grouping comprises
the optically and X-ray luminous systems Wd1-36 and -53a. A further ten optically faint 
O+O binaries of uncertain spectral type and luminosity class are excluded from the table, as is the sole Be star; 
none of these are X-ray detections.}
\end{table}

\section{Motivation, methodology \& results}

The above empirical results derive from surveys of ensembles of field stars (e.g. Bergh\"{o}fer et al. \cite{bergh}, Oskinova \cite{oskinova05}, Nebot G\'{o}mez-Mor\'{a}n \& Oskinova \cite{nebot}) and targeted observations of star clusters and star forming regions (Sana et al. \cite{sana}, Naz\'{e} et al. \cite{naze11}, Rauw et al. \cite{rauw15}). The former benefits from large  sample sizes that span a range of spectral types but fundamental stellar properties and evolutionary status can be difficult to determine. Conversely, while the distance to, and ages of, stars within clusters can be well constrained, sample sizes are typically relatively small; this may be alleviated by studying OB associations, although such aggregates demonstrate  complex, extended star formation histories. 

The young, massive cluster Westerlund 1 offers a uniquely rich and co-eval population of massive stars which, unlike galactic centre clusters such as the Quintuplet and Arches, may be readily resolved by extant X-ray instrumentation (Clark et al. \cite{clark05}, Negueruela et al. \cite{iggy10}, Kudryavtseva et al. \cite{kud}). Moreover, it has been  identified as the source of  highly  energetic $\gamma$-rays (GeV and TeV; Ohm et al. \cite{ohm} and Abramowski et al. \cite{abramowski}, respectively) and consequently has  been implicated in the production of cosmic rays (e.g. Bykov et al. \cite{bykov}, Aharonian et al. \cite{aharonian}), providing a powerful motivation for  understanding of the nature of the X-ray emitting stellar cohort.

Clark et al. (\cite{clark08}) found a total of 53 X-ray sources that appeared to be associated with the massive stars and presented spectral classifications for half of these. However, since this study, classifications have become available for 166 cluster 
members (cf. Negueruela et al. \cite{iggy10}, Clark et al. in prep.) and so it seems opportune to revisit this dataset. In total we are now able to provide classifications for 45 X-ray detections, in part comprising 19 new objects and reclassification of a further 13 stars. Moreover the binary status of 22 X-ray bright cluster members have been investigated  via our VLT/FLAMES radial velocity (RV) survey (Ritchie et al. \cite{ritchie09}, \cite{ritchie10b}, Clark et al. in prep.), with  VLT/UVES observations available for two further stars (Wd1-27a and WR L; Clark et al. \cite{clark18} and in prep.); relevant data reduction and analysis techniques are described in these works. 

 These yield binary RV periods for Wd1-13, WR  F and L, while Bonanos (\cite{bonanos}) reports photometric periods 
for Wd1-6a, -36, -53 and WR A and B\footnote{For reasons of space and legibility we have abbreviated the Simbad recognised nomenclature; the formal forms being Cl* Westerlund 1 W{\em xx} for those stars listed in Westerlund (\cite{westerlund}) and Cl* Westerlund 1 CN {\em x} for the Wolf-Rayet cohort. Star designated W1{\em xxx} are new cluster members presented in Clark et al. (in prep.) and are ordered by increasing right ascension.}; these eight systems therefore appear to be unambiguous binaries (Table A.1). The optical spectrum of 
Wd1-10 in Clark et al.  (\cite{clark08}) demonstrates a double-troughed He\,{\sc i} 7065{\AA} photospheric line that is 
indicative of a SB2 classification  and  one epoch of our  $I-$band  observations clearly confirms this hypothesis (Fig. 
1). Intriguingly, in the remaining epochs  Wd1-10 has the appearance of a single star, suggesting that it may be a highly 
eccentric system with the double lined spectrum  fortuitously obtained at periastron. For the purpose of this paper we therefore  consider it a {\em bona fide} binary (cf. Fig. 2).

Wd1-24, -30a and W1040 are found to be  RV variable at a level indicative of binarity, although no unique period may be identified for any object in the current datasets (Clark et al. \cite{clark10}, \cite{clark18} and in prep.). Additional variability is observed in the C\,{\sc iii} 8500{\AA}/Pa-16 blend of the first two systems, suggesting the presence of a second O star of  similar luminosity (Clark et al. \cite{clark10}).
W1041 demonstrates very broad, flat-bottomed photospheric Paschen-series lines that are indicative of an 
unresolved blend of two  stars of comparable luminosities, an hypothesis bolstered by the morphologically identical spectra of the photometrically confirmed binaries Wd1-36 and -53a  (Fig.  1; Bonanos \cite{bonanos}). We therefore conclude that  all four stars are compelling binary candidates, although not yet unambiguous classifications in the manner of the preceding nine systems. 

Several authors have suggested that supergiant B[e] stars are massive (post-) interacting binaries (Kastner et al. \cite{kastner}, Clark et al. \cite{clark13a}) and two additional lines of evidence suggest that this is the case for Wd1-9. Specifically radio- and mm-continuum observations reveal the presence of a bipolar outflow with a velocity and mass-loss rate that is inconsistent with expectations for a radiatively driven  wind from a single supergiant or Wolf-Rayet (Fenech et al. \cite{fenech17}, \cite{fenech18}). However the mass-loss rate is comparable to predictions for mass-transfer from the primary in a compact binary  undergoing rapid case-A evolution (Petrovic et al. \cite{petrovic}). Likewise, the presence of forbidden  [O\,{\sc iv}] emission  in the mid-IR spectrum of Wd1-9 requires either stellar temperatures in excess of those expected for  single OB supergiants  ($\gtrsim60$kK; Clark et al. \cite{clark98}, \cite{clark13}) or shock excitation, which would naturally arise in the  WCZ of a massive interacting binary. The non-thermal radio continuum spectrum of Wd1-17 and  
the association of hot dust with the WC9 star WR N are likewise indicative of binarity for these stars (Clark et al. \cite{clark08}, Dougherty et al. \cite{dougherty}). Hence we conclude that  these three cluster members are likely to be  binaries, yielding a total of 16 confirmed and candidate binaries associated with X-ray emission.

Finally we briefly consider Wd1-65 and -232. Wd1-65 (O9 Ib) appears to be a low level RV variable with an I-band spectrum
suggestive of binarity; while the Paschen lines are typical of lower-luminosity O9 stars, the C~III/Pa-16 blend appears anomalously broad and weak (cf. Fig. 1). Likewise, the star seems rather too luminous for its implied early spectral type (Clark et al. in prep.). Erroneously listed as W234 in Ritchie et al. (\cite{ritchie09}) Wd1-232 (O9.5 Ib) is a clear RV variable, albeit of low amplitude. As a consequence period fitting failed to yield a  unique value, with multiple possible short periods ($<10$days) returned.  However, significant scatter is present in the RV curve when folded onto any of these periods and hence the nature of this object remains unclear, with either complex pulsational modes or a combination of pulsational and orbital
periodicities possible explanations for the RV variability (Ritchie et al. \cite{ritchie09}, Clark et al. in prep.). We therefore 
refrain from listing these stars as strong binary candidates  at this time, pending further observations.

We present the new and revised spectral classifications  for the 45 X-ray sources in Table A.1, along with information on 
binarity and basic X-ray properties. Following Clark et al. (\cite{clark08}), Fig. 2 summarises the spectral properties and fluxes of these sources as a function of spectral type and binarity. Using the Portable Multi-Mission Simulator
(PIMMS)\footnote{http://heasarch.gsfc.nasa.gov/Tools/w3pimms.html} and XSPEC, we simulated the colours and fluxes produced by a thermal plasma spectrum absorbed by a column equivalent to $2\times10^{22}$ cm$^{-2}$
of H - appropriate for the mean extinction to the cluster (Negueruela et al. \cite{iggy10}), following the relationship of
Predehl \& Schmitt (\cite{PS}) - for a range of  temperatures and intrinsic luminosities (0.5--8.0 keV, assuming a distance of 5kpc). The resulting grid is overplotted on Fig. 2 to aid in the interpretation of individual sources. Finally Table 1 summarises the stellar demographics of the X-ray emitting cluster population.

\section{The nature of the emitters}
Table 1 clearly summarises the first of two major findings of this work; X-ray emitters are found to be confined to very limited subsets of cluster members. The lack of emission amongst the cool stellar cohort had already been recognised (Clark et al. \cite{clark08}), but despite the increase in numbers of B5-9 Ia$^+$ and B2-4 Ia stars (11 versus 6) and the identification of ten new B1-1.5 Ia,Iab stars, none are detected; consistent with the findings of e.g. Cassinelli et al. (\cite{cassinelli}) and Bergh\"{o}fer et al. (\cite{bergh})\footnote{Cazorla et al. (\cite{cazorla}) and Oskinova et al. (\cite{oskinova17}) report strong X-ray emission for the B3-4 Ia$^+$ star Cyg OB2 \#12 (Clark et al. \cite{clark12}), which they attribute to putative binarity. None of the six B3-9 super-/hypergiants within Wd1 are found to be X-ray detections, nor do any show signs of binarity.}. 
Instead  emission appears concentrated amongst the O9-O9.5 Iab,Ib and B0-0.5 Ia,Iab,Ib supergiants (25 from 53) with earlier spectral types within this range clearly favoured (Table 1). Of the 39 lower luminosity O9-9.5 II,II-III,III stars only W1033 (O9-9.5 I-III) and W1050 (O9 III) are detected (with the former having an uncertain luminosity class). The remaining 
X-ray emitters appear to comprise a number of interacting/post-interaction OB+OB binaries and  a subset of the cluster WRs; 
we anticipate a significant overlap between these latter cohorts, in the sense that a number of WRs are likely post-interaction systems.

\begin{figure*}
\begin{center}
\includegraphics[width=12cm,angle=-90]{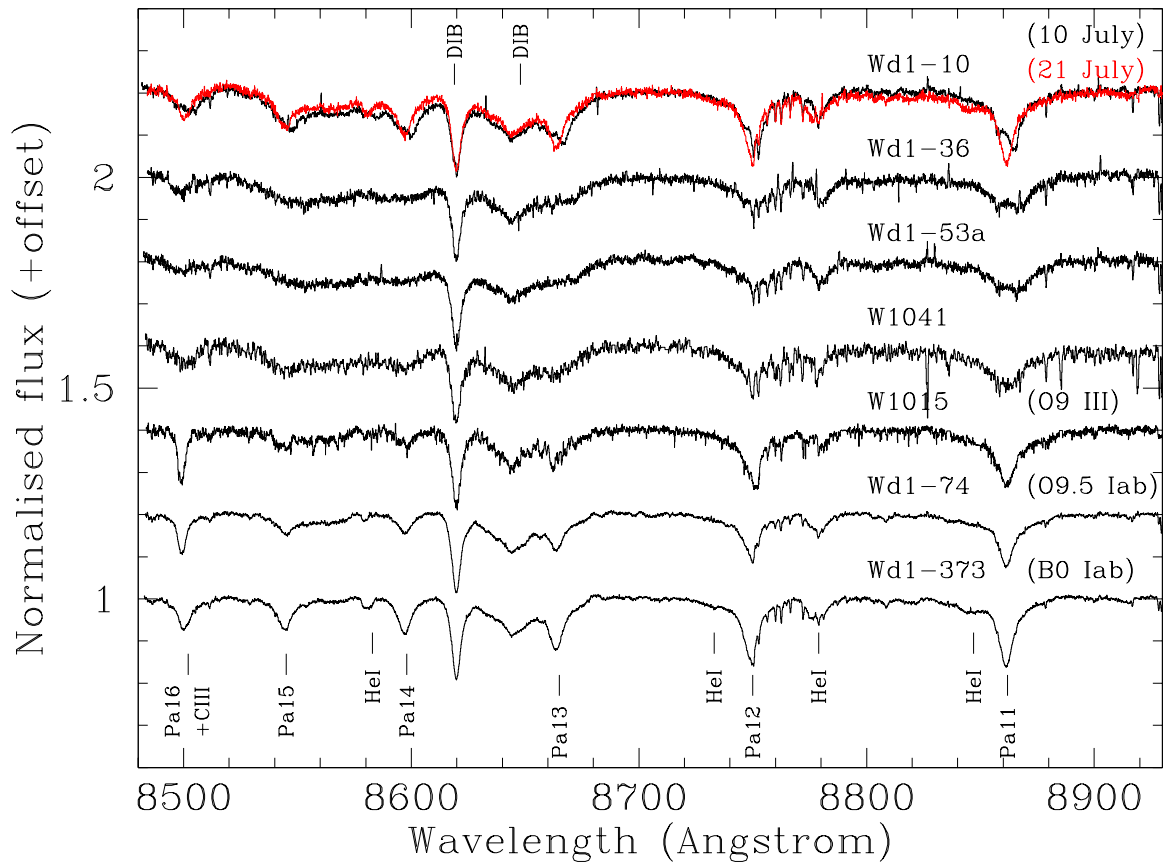}
\caption{Montage of classification  spectra (Wd1-74, -373 and W1015) and apparent SB2 binaries. Note the shallower, 
flat-bottomed photospheric profiles for Wd1-36, 53a and W1041 in comparison to the templates, indicative of unresolved SB2 
systems.
Two spectra
are presented for Wd1-10 to demonstrate the pronounced line profile variability in the Paschen-11 and -12 photospheric lines, which transition from  doubled- to single-troughed in the 11 days between the observations.} 
\label{fig:standards}
\end{center}
\end{figure*}  

\begin{figure*}
\begin{center}
\includegraphics[width=15cm,angle=-0]{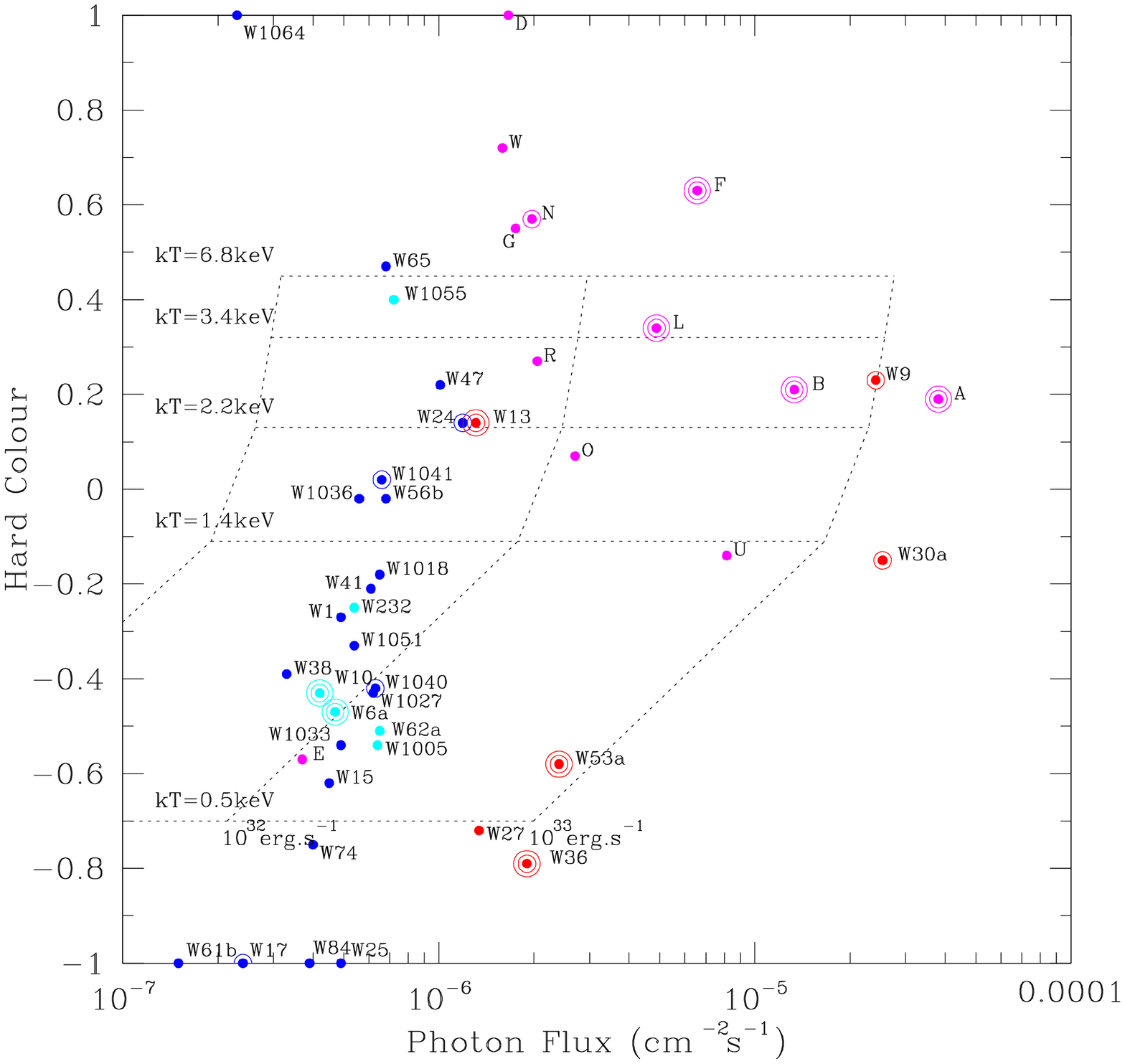}
\caption{Hardness-intensity diagram illustrating the spectral properties as a function of intensity for the X-ray point sources within the 5'x5' field centred on Wd~1. The hardness colour  is defined as (h-s)/(h+s) where h are the counts in the 
2.0-8.0 keV band and s are the counts in the 0.5-2.0 keV band. Errorbars are not shown for reasons of clarity but are inversely correlated to flux and are given for individual sources in Table A.1. Two concentric rings denote comfirmed binaries, while one ring designates strong candidate systems (see Sect. 2).
 Blue and cyan symbols  are O9-9.5 Iab,Ib,II,III and B0-0.5 Ia,Iab,Ib supergiants respectively, purple symbols are WRs and red symbols correspond to SB2 binaries of uncertain classification (Wd1-36 and  -53a), binary products (Wd1-27 and -30a) the  interacting binary Wd1-9 and eclipsing WNVL/BHG+OB system Wd1-13. Note that following the discussion in Sects. 2 and 3.2, Wd1-24 would also be included in this latter grouping if its binary nature were confirmed.}
\label{fig:standards}
\end{center}
\end{figure*}

\subsection{O9-B0.5 supergiants} 
We expect {\em individual} OB supergiants within Wd~1 to have  
$\sim30-40M_{\odot}$ progenitors and hence $L_{\rm bol}\sim
3-5\times10^5L_{\odot}$ after $\sim5$Myr (Clark et al. \cite{clark05}, 
Negueruela et al. \cite{iggy10}, Ritchie et al. \cite{ritchie10a}). Assuming 
L$_{\rm X} \sim 10^{-7}$L$_{\rm bol}$ this would 
imply L$_{\rm X} \sim 1.2-2\times10^{32}$erg s$^{-1}$.
The OB supergiant detections have X-ray fluxes broadly consistent with 
this prediction (Fig. 2)\footnote{Unfortunately, the 90\% completeness limit for a 0.6keV emitter is  
2$\times 10^{32}$erg s$^{-1}$ (assuming $d$=5 kpc and $N_{\rm H} =
2\times10^{22}$ cm$^{-2}$), hence we are likely incomplete for
emission from single OB supergiant stars}. Due to the low count rates, uncertainties on the hardness
colour are large for photon fluxes $<10^{-6}$cm$^{-2}$ s$^{-1}$.
Nevertheless, eight of the 27 OB SG detections have a hardness colour deviating 
from $\sim-0.5$ by $\geq1\sigma$; appropriate for
shock-heated material within single O star winds ($\sim0.6$keV; Feldmeier et
al. \cite{feld}).

 An obvious explanation for these harder sources would be emission from hot shocked gas in a WCZ in a massive binary (but 
see below).  Of the eight, Wd1-24 and W1041 are strong binary candidates, with tentative evidence for binarity for Wd1-65. 
W1036 and W1055  show no indication for binarity in our RV data, while Wd1-47, -56b and W1064 have yet to be subject to RV monitoring, but are not SB2s. If future observations were 
to  confirm binarity for Wd1-65 and the latter three objects, one would suspect that W1036 and W1055 are also binaries seen 
at an unfavourable inclination for identification via RV shifts. 

Given their less evolved status, and hence lower bolometric luminosities, one might suppose that  both W1051 (O9 III) and 
W1033 (O9-9.5 I-III) might also be (X-ray overluminous) binaries, but no evidence for this has been found in  our RV data. 
While we cannot rule out binarity, Nebot G\'{o}mez-Mor\'{a}n  \& Oskinova (\cite{nebot}) find a large scatter  in the 
$L_{\rm X}/L_{\rm bol}$ relation and, if their  classifications are correct, these two stars may simply be the  brightest 
examples of their evolutionary phase within Wd1.

 It is notable that  a number of binaries with primaries spanning spectral types O9-B4 and luminosity class I-III - e.g.
Wd1-43a (B0 Ia +? and $P_{\rm orb}\sim16.27$d; Ritchie et al. \cite{ritchie10b}) and  Wd1-52 
(B1.5Ia +? and $P_{\rm orb}\sim6.7$d; Bonanos \cite{bonanos}) -  are not detected as X-ray  sources. Moreover,
 binaries such as Wd1-6a (B0.5 Iab +? and $P_{\rm orb}\sim2.20$d; Bonanos \cite{bonanos}) and Wd1-10 (B0.5 I +OB; Negueruela et al. \cite{iggy10}) have X-ray luminosities and hardness colours that are  directly comparable to those of  apparently single stars of similar spectral type (e.g. Wd1-62a (B0.5 Ib) and W1005 (B0 Iab); Fig 2).
These findings are consistent with a number of recent observational studies which  indicate that, contrary to historical expectations, the majority of binaries  do not generate excess hard X-ray emission (e.g. Oskinova \cite{oskinova05}, Naz\'{e} \cite{naze09}, Naz\'{e} et al. \cite{naze09}, Rauw et al. \cite{rauw15},
Nebot G\'{o}mez-Mor\'{a}n  \& Oskinova \cite{nebot}).  Instead, they follow the same canonical $L_{\rm X} \sim 10^{-7}L_{\rm bol}$ relation as single stars, suggesting that the bulk of the emission arises in the stellar winds of the individual stars, rather than in  WCZs.

The lack of X-ray detections amongst the  non-supergiant O-star binary population of Wd1 (Clark et al. in prep.) would be consistent with this hypothesis; presumably both components in such systems support rather weak winds in comparison to the supergiant cohort, such that they are neither intrinsically X-ray luminous nor can generate excess emission in a WCZ. 
 Similarly, an obvious explanation for the absence of spectral features from the secondary in SB1 binaries such as Wd1-6a, 43a and  -52 is that they are less luminous and evolved stars supporting comparatively weak winds. Consequently a WCZ is either weak or absent, resulting in the detectability of the system being solely dependent on the nature of the primary. This would naturally explain the similarity of Wd1-6a to X-ray bright  single O supergiants such as Wd1-38 and -62a (see also Sect. 4.2), while the lack of detections for Wd1-43a and -52 (and  single stars of comparable spectral type and luminosity class) would result from a combination of the
 significant intrinsic scatter in the $L_{\rm X}/L_{\rm bol}$ relation (Nebot G\'{o}mez-Mor\'{a}n  \& Oskinova \cite{nebot})
and incompleteness (cf. footnote 4). Clearly, deeper observations would help resolve this issue.  

However, the properties of Wd1-10 are more challenging to interpret, given that the presence of double troughed He\,{\sc i} and Paschen series lines  (Fig 1; Negueruela et al. \cite{iggy10}) clearly requires the secondary to be of comparable luminosity to the B0.5 I primary.  As such it appears similar to a number of other SB2 binaries within Wd1 (e.g. Wd1-13, -24, -36 and -53a); however these are brighter and/or harder X-ray sources than Wd1-10 (Sect. 3.2). The most obvious explaination for this discrepancy is that Wd1-10 is a highly eccentric system (cf. Sect. 2) yielding an orbital  dependence for X-ray luminosity and spectral shape and that our X-ray observations happened to catch the binary at a phase where emission from a WCZ is minimal/absent.

\subsection{Interacting and post-interaction OB binaries}
As summarised in Rauw \& Naz\'{e} (\cite{rauw}), the weight of observational evidence increasingly suggests that the X-ray properties of many massive binaries do not deviate from those of single stars. However this conclusion does not hold for {\em all} binaries, with $\eta$ Carina being an exemplar in which orbital modulation of the WCZ leads to substantial variability in the  X-ray spectrum and overal luminosity  (e.g. Corcoran et al. \cite{corcoran} and refs. therein). Indeed
Wd1 contains a number of massive interacting and post-interaction systems that are found to be anomalously hard and/or luminous in comparison to the late-O/early-B supergiants discussed in the preceding section (Fig. 2).

Of these, both Wd1-9 and -30a are amongst the subset of sources that have enough counts for tailored spectral analyses; Clark et al. (\cite{clark08}) report fluxes of $\sim3.6\times10^{33}$ergs$^{-1}$ and
$\sim1.6\times10^{33}$ergs$^{-1}$ respectively and, critically,  harder spectra than expected for single stars ($kT\sim2.3^{+0.5}_{-0.3}$ and
 $\sim1.3^{+0.1}_{-0.1}$). Individual absorption columns were determined for  both stars, contrasting with the single global value 
adopted for all cluster members following the methodology outlined in Sect 2; this  leads to a minor discrepancy between the flux estimates for Wd1-9 determined by the two approaches (cf.  Fig. 2). Any such evaluation is complicated by the fact that significant differential extinction exists across the cluster and the form of the reddening law is uncertain (Clark et al. \cite{clark18}). 
Given this uncertainty and the quality of the X-ray spectra, re-analysis of extant  data appears premature. 

However the overal conclusion - that these stars appear anomalously bright and hard relative to the majority of cluster OB supergiants - remains valid under either methodology, even if the absolute flux is uncertain by a factor of a few. Indeed, both Wd1-9 and -30a have X-ray fluxes that would place them amongst the   most  luminous 20\% of known O+O binaries (Gagn\'{e} et al. \cite{gagne})\footnote{We caution that the majority of such binaries consist of two mid-late O dwarfs and hence may be expected to be intrinsically faint.}, with Wd1-9 $\gtrsim6\times$ times brighter than expected for canonical OB supergiants within Wd1. 

Since the stellar components of Wd1-9 are entirely veiled by circumstellar material (Clark et al. \cite{clark13}) there has been some ambiguity regarding the nature of  the system -  colliding wind or accreting binary. Recent observations of the sgB[e] X-ray binaries CI Cam and NGC300 ULX-1 in quiescence reveal a power-law component due to accretion onto a compact companion 
(Bartlett et al. \cite{liz}, Carpano et al. \cite{carpano}). Such a spectral component is absent in Wd1-9, which instead closely resembles  confirmed colliding-wind binaries  such as WR A and B (and Wd1-30a). Given this, we consider it most likely that Wd1-9 is a massive binary undergoing rapid, case A mass transfer (cf. Sect. 2).

Unlike Wd1-9, detailed analysis of the dominant visible component of Wd1-30a is possible, and suggests that it is the massive, rejuvenated secondary in a post-interaction binary (Clark et al. \cite{clark18}). With a spectral type of O4-5 Ia$^+$ it is the hottest OB star within Wd1 and has a luminosity of log$(L_{\rm bol}/L_{\odot})\sim5.87^{+0.15}_{-0.10}$, making it a factor of $\gtrsim5\times$ more X-ray luminous than expected for a single star. Indeed the hardness of the emission implies an additional (dominant) contribution from a WCZ, with RV data indicating it still resides in a binary with a likely period of $\lesssim10$ days
 (Clark et al. \cite{clark18}). 

A number of binaries with mid-O super-/hypergiant `primaries' and X-ray luminosities $\geq10^{32}$ergs$^{-1}$ 
are reported by Gagn\'{e} et al. (\cite{gagne}) and Cazorla et al. (\cite{cazorla})\footnote{
HD~93403 (O5.5 I + O7 V, $P_{\rm orb}\sim15.093$d),
HD~47129 (O7.5 I + O6 I, $P_{\rm orb}\sim14.4$d),
Cyg OB2 \#5 (O6.5-7 I + Ofpe/WN9, $P_{\rm orb}\sim6.6$d),
Cyg OB2 \#8A (O6 If + O5.5 III(f), $P_{\rm orb}\sim21.9$d) and 
Cyg OB2 \#9 (O5 If + O6-7, $P_{\rm orb}\sim852.9$d)}. 
Wd1-30a currently presents as an SB1 system, but one might expect the original primary to be an  H-depleted late WNLh star
similar to Wd1-5 (Clark et al. \cite{clark14}); as a consequence its apparent similarity to Cyg OB2 \#5 in terms of composition, possible orbital period and X-ray luminosity is intriguing. However Cazorla et al. (\cite{cazorla}) suggest that the X-ray emission in Cyg OB2 \#5  varies on the 6.7yr periodicity of a tertiary late-O/early-B component (Kennedy et al. \cite{kennedy}) and consequently may not originate in the close binary component. 

Instead, the best comparators to Wd1-30a  may be the less evolved/extreme (post-) interacting binary Wd1-13 -  comprising an H-depleted  B0.5 Ia$^+$/WN10h primary and mass gaining OB Ia secondary ($P_{\rm orb}\sim9.27$d; Ritchie et al. \cite{ritchie10a}) - and, subject to confirmation of binarity, its  X-ray and potential evolutionary twin  Wd1-24 (see Fig. 2 and Sect. 2).  Despite being an order of magnitude fainter than Wd1-30a, both demonstrate   similarly hard X-ray emission, suggesting  that a contribution  from a WCZ dominates their spectra. If this comparison is correct, one might speculate that the stronger wind of the mass-gaining secondary in Wd1-30a leads to the difference in flux if all three binaries are otherwise physically similar. Alternatively the initial primary in Wd1-30a may have evolved further to a become an H-free WN or WC star. Given the  rarity of mid-O hypergiants and that no trace of a WR companion 
would be expected in optical spectroscopy we cannot identify a compator system for such a putative configuration, but WR binaries with less extreme companions than O4-5 Ia$^+$ routinely reach such X-ray luminosities (Gagn\'{e} et al. \cite{gagne}; Sect. 3.3).

Finally we turn to Wd1-27, -36 and -53a, which are co-located on the X-ray flux/hardness plot (Fig. 2) with soft spectra but
 high luminosities ($L_{\rm X}\sim10^{33}$ergs$^{-1}$); as with Wd1-9 and -30a they are amongst the most luminous OB+OB binaries known (Gagn\'{e} et al. \cite{gagne}). Bonanos (\cite{bonanos}) identifies the latter two objects as short period eclipsing and elipsoidally  modulated systems respectively; our spectroscopy reveals that both are SB2 binaries comprising twin  highly luminous and evolved OB stars (Fig. 1). Detailed classifications of the components are  not possible due to the blending of the Paschen series lines but the absence of cool N\,{\sc i} photospheric lines precludes a spectral type of B2 or later and by comparison to other spectra we suspect they are substantially earlier. 
By contrast Wd1-27 is similar to Wd1-30a, showing evidence for substantial mass accretion in a binary, leading to a classification of O7-8 Ia$^+$ and log$(L_{\rm bol} / L_{\odot})\sim5.97^{+0.15}_{-0.10}$ (Clark et al. \cite{clark18}). No RV motion has been detected for this object; hence we cannot distinguish between e.g. a binary with an extreme current mass ratio observed under an unfavourable inclination or the product of binary merger (Clark et al. \cite{clark18}). 

It is therefore of interest that the soft  X-ray emission of Wd1-27 is consistent with expectations for a single star, while the flux is only $\sim2\times$ greater than one would predict for a star of its extreme luminosity. Bergh\"{o}fer \&  Schmitt (\cite{B95}) analysed the short period ($P_{\rm orb}\sim4.39$d)  O8 Iaf + O9 I binary UW CMa, finding the X-ray emission arose from a summation of the individual contributions from the stellar winds of both components, with no need to invoke a WCZ. Given the apparent similarity of UW CMa to Wd1-36 and -53a and after consideration of Wd1-27,   one might wonder whether a similar scenario could also account for  their X-ray properties.

To summarise; the multiwavelength properties of Wd1-9, -13, -24 and -30a reveal them all to be massive, evolved and interacting binaries, with their hard spectra revealing that emission from  WCZs likely dominates their X-ray flux. Conversely, despite Wd1-36 and -53a being short period systems and the potential for (current) binarity for Wd1-27, we find no unambiguous
evidence for emission from a WCZ in these systems; we discuss these latter systems further in Sect. 4.2.

\subsection{Wolf-Rayets}

Oskinova (\cite{oskinova16}) provides a review of the observational properties of WRs, reporting that they are diverse X-ray emitters, with fluxes ranging over several orders of magnitude.  Broken down by subtype, single WNL stars are relatively weak emitters ($L_{\rm X}\sim10^{32}$ergs$^{-1}$; Oskinova \cite{oskinova05}) with WNE stars found to be more luminous
($L_{\rm X}\sim2-6\times10^{32}$ergs$^{-1}$; Oskinova \cite{oskinova16b}). Likely due   to the opacity of their dense, metal-rich winds, WC stars are significantly fainter, with only one example of an apparently single star, WR144 (WC4),
detected to date  ($L_{\rm X}\sim10^{30}$ergs$^{-1}$; Rauw et al. \cite{rauw}). Intriguingly, single WR stars do not conform to the  $L_{\rm X}/L_{\rm bol}$ relation of O stars and, on  average, binaries are found to be brighter (Pollock \cite{pollock}), with fluxes ranging up to $L_{\rm X}\sim10^{35}$ergs$^{-1}$ (the WC8d star WR48a; Zhekov et al. \cite{zhekov}).

The X-ray properties of the WR population of Wd1 appear consonant with these expectations.  The distribution of the spectral subtypes of the X-ray emitters is formally identical to the underlying distribution they are drawn from (Table 1; Clark et al. \cite{clark08}). As can be seen from Fig. 2, with the  notable exception of WR E, detected systems are on average harder and/or more luminous than the OB supergiant cohort, overlapping the region of the hardness/flux diagram that the  interacting OB supergiants occupy (Sect. 3.2)\footnote{Upon individual modeling, the absorption columns found for WR A and  B (WR L) 
imply an excessively high (low) visual extinction in comparison to optical determinations (Negueruela et al. \cite{iggy10}), resulting in a likely  over- (under-) estimate of the relevant  X-ray fluxes determined via this methodology (Clark et al. \cite{clark08} and Sect. 3.2).}. 
The brightest stars, WR A and WR B are long-known compact binaries with WN7 primaries ($P_{\rm orb}\sim7.63$d and $\sim3.51$d respectively; Bonanos \cite{bonanos}). Our VLT/FLAMES data shows the WC9d star WR F to be a similar short-period binary ($P_{\rm orb}\sim5.05$d), with long term IR variability suggesting the presence of a  tertiary component (Clark et al. \cite{clark11}); as a consequence it is not obvious which components contribute to the WCZ responsible for the X-ray emission. Recent RV monitoring reveals the WN9h star WR L (=Wd1-44) to be a longer period  binary ($P_{\rm orb}\sim54$d; Clark et al. in prep.); as such four of the five  brightest and hardest X-ray detected WRs within Wd1 are unambiguously confirmed to be massive binaries (cf. Skinner et al. \cite{skinner}).

 Unfortunately, due to the combination of high interstellar extinction towards Wd1 and the capabilities of VLT/FLAMES, a lack of appropriate spectral diagnostics prevents an RV survey for binarity amongst the WN5-8 stars within Wd1. Nevertheless we strongly suspect binarity for WR U given its luminosity and spectral similarity to WR A and B (Clark et al. \cite{clark08}), while the hard but fainter sources WR D (WN7o), G (WN7o) O (WN6o), R (WN5o) and W (WN6h) all appear strong binary candidates. The dusty nature of the hard but comparatively X-ray faint WC9d WR N also implies binarity (cf. Williams et al. \cite{williams}). The binary status  of the WC9 star WR E is unclear at this time but it is unique amongst the WRs as being a rather faint and soft X-ray source co-located with the OB supergiants in the hardness/flux diagram (Fig. 2).

While the binary nature of the remaining undetected WN stars is uncertain,  the presence of an IR-excess due to hot dust in the  WCL  stars WR C, H and T implies that they are binaries. Since binary induced dust production and X-ray emission in WRs  modulate on the orbital period, one might suppose that these stars were observed at an unfavourable  phase.

\subsection{Marginal detections}
 Finally we discuss the handful of massive cluster members that are found in a post-supergiant evolutionary phase and are marginal X-ray detections at the $\sim90$\% level of 
significance (Clark et al. \cite{clark08}). Assuming a 0.5keV thermal plasma appropriate for a single early-type star, such sources have fluxes $\sim10^{31}-10^{32}$ergs$^{-1}$; as such WR Q (WN6o), S (=Wd1-5; WN10h), V (WN8o) and X (WN5o) would appear to represent the low brightness tail of the luminosity distribution observed for WRs (Sect. 3.3). Of these WR S is of interest since it appears to be the product of binary-induced mass-stripping although, presumably following the explosion of its putative companion,  it now appears to be single (Clark et al. \cite{clark14}). In terms of gross properties WR S appears similar to the primaries of Wd1-13 and WR L; both of which are hard and luminous X-ray sources (Fig. 2 and Table A.1). As such it appears likely the presence of a WCZ in the latter two system is the cause of the discrepant X-ray properties.
 
Surprisingly, the last two objects for which tentative detections were made are the luminous blue variable (LBV) Wd1-243 and the yellow hypergiant Wd1-8a, with the former in a cool phase when the X-ray observations were undertaken (spectral type $\sim$A3 Ia$^+$; Clark \& Negueruela \cite{clark04}, Ritchie et al. \cite{ritchie09b}). 
Naz\'{e} et al. (\cite{naze12}) investigated the X-ray properties of LBVs, finding emission to be rare and potentially associated with binarity. With the X-ray flux of Wd1-243 being fully consistent with cluster OB supergiants, emission from an unseen companion would seem an obvious explantion for its properties. Indeed, the  strong H\,{\sc i} and He\,{\sc i} emission lines in the spectrum of Wd1-243 leads 
Ritchie et al. (\cite{ritchie09b}) to infer the presence of a hot, luminous companion, since the temperature of the LBV `primary' is 
insufficient to ionise its wind.

A similar explanation would appear appropriate for Wd1-8a although, since it is located within the crowded core of Wd1, it is not obvious if it is a {\em bona fide} binary or instead if the putative OB supergiant is simply located along the same line-of-sight.

\section{Interpretation and concluding remarks}
Significant progress has been in the spectral classification of cluster members and the determination of their binary status in the decade since the publication of the X-ray point source catalog. These new data allow a more accurate and concise characterisation
of the nature of the massive X-ray emitting population of Wd1.  We may draw two fundamental conclusions from this; that X-ray  emitters are confined to very well delineated stellar subsets and that binarity plays an important role in the emission properties of  some, but not all, of these stars.

\subsection{The influence of luminosity and spectral type}
X-ray emission is confined to three subsets of stars: WRs, interacting binaries and O9-B0.5 Ia,ab,b supergiants  and, of the latter cohort, it is biased towards earlier (O9-9.5) spectral types (Table 1). The properties of the WRs appear broadly consistent with empirical expectations (Sect. 3.3) and hence we defer discussion of these and the (post-) interacting OB binaries until later, concentrating here on the (single) OB supergiants. 
The lack of emission from the less evolved O9-9.5 II-III stars is most likely due to their comparatively low bolometric luminosities, since they  are expected 
to follow the $L_{\rm X}/L_{\rm bol}$ relation (cf. Nebot G\'{o}mez-Mor\'{a}n  \& Oskinova \cite{nebot}). However, one cannot appeal to  a low bolometric luminosity to account for the absence of emission in super-/hypergiants of spectral type B1 and later, since  supergiants  of $M_{initial}\sim30-40M_{\odot}$ should evolve to cooler tempertures at {\em constant}  luminosity (e.g. Ekstr\"{o}m et al. \cite{ekstrom}). As a consequence  if the $L_{\rm X}/L_{\rm bol}$ relation were to apply for {\em all} B supergiants, we would not expect emission within Wd1 to abruptly cease at spectral type B0.5.  

Our results mirror those of Bergh\"{o}fer et al. (\cite{bergh}), who find X-ray emission to abate at spectral type B1 for stars of  luminosity class I and II. While the physics resulting in the  $L_{\rm X}/L_{\rm bol}$ relation is uncertain, 
Bergh\"{o}fer et al. (\cite{bergh}) suggest that beyond this point the wind velocity drops to such a degree that it is too low  to produce the strong shocks required to yield X-ray emission. Prinja et al. (\cite{prinja}) and Lamers et al. (\cite{lamers})  report a discontinuity in wind properties around spectral type B1, with winds of high (low) velocity and low (high) mass-loss rate above (below) this point. Vink et al. (\cite{vink}) attribute this discontinuity to a change in the ionisation balance of iron in the wind, which modifies the efficiency of line driving. Intriguingly, circumstantial evidence for such a  bistability
jump is seen in mm-continuum observations of Wd1, where stars are only detected at spectral types B0 and later; on the cool side of this transition. Since thermal continuum emission from the stellar wind is sensitive to its density (scaling  as (\.{M}/$v_{\rm wind}$)$^{4/3}$; Wright \& Barlow \cite{wright}), it is tempting to attribute this finding to an abrupt increase in this parameter for such stars  (cf. Fenech et al. \cite{fenech18}). 

One might therefore suppose that  at temperatures above the bistability jump, wind velocities are high enough to permit X-ray emission via wind-shocks while, at temperatures  below this transition, wind densities are sufficient  to yield strong thermal continuum emission, but velocities are insufficient to generate post-shock temperatures high enough to produce X-ray emission.

\subsection{The influence of binarity}

As implied  by Rauw \& Naz\'{e} (\cite{rauw}), the role of binarity in determining the properties of  the X-ray emission from cluster members appears complex. Clark et al. (in prep.) infer the presence of a large number of massive binaries within Wd1, but many are not detected at X-ray energies. This result is consistent with recent studies that suggest that in many binaries the WCZ, if present, does not yield excess, hard X-ray emission (Sect. 3.1). This behaviour is observed  for the subset of binaries containing luminous OB star primaries within Wd1, with the X-ray properties of   e.g. Wd1-6a and -10 being indistinguishable from those of single supergiants of comparable spectral type. Nevertheless, emission from a WCZ clearly plays a role in a subset of cluster binaries  which are both harder  and brighter than expected for single stars. Four of the five brightest and hardest WR X-ray detections are unambiguously binary (WR A, B, F and L), while the interacting and post-interaction systems Wd1-9 and -30a have emission that also appears to be  dominated by a WCZ. A number of fainter WR and O supergiant detections also appear anomalously hard, including the confirmed short-period binary Wd1-13. 

However, while the the short period OB+OB binaries Wd1-36 and -53a are also anomalously bright, they are unexpectedly soft; properties  that are potentially consistent  with the summed emission from the stellar components with no contribution from a WCZ, {\em despite} their apparent similarity to e.g. Wd1-13 (see also Wd1-24; Sect. 3.2).  Notwithstanding this finding, we may conclude that while binarity does not lead to  {\em all} binaries being more luminous than single stars, the hardest and most luminous stellar X-ray sources within Wd1 are either interacting binaries comprising stars of high bolometric luminosity or post-interaction systems containing either a WR or rejuventated binary product (cf. Wd1-30a).

Unfortunately interpreting this diverse behaviour is difficult. The contribution of a WCZ to the intrinsic emission of the binary components will depend on the  cooling efficiency for shocked material and the attributes of both outflows at the location of the interaction (Stevens et al. \cite{stevens}). The properties  of the stellar winds in turn depend on the nature of the driving stars as well as binary separation and eccentricity (i.e. will the winds have time to reach their terminal velocity and is radiative braking important?). While beyond the scope of this paper,  tailored hydrodynamical simulations will be required to fully understand the physical effects leading to the manifold X-ray properties of apparently similar binary systems (e.g. Pittard \& Dawson \cite{pittard}). 

Nevertheless, it is of interest to consider Wd1-6a, -36 and 53a, for which emission from a WCZ appears absent and photometric periods are available. Following the analysis of Wd1-13 by Ritchie et al. (\cite{ritchie10a}) we adopt a current total mass of $\sim60M_{\odot}$ for such 
systems and stellar radii for the constituent  late O/early B supergiants of $\gtrsim20R_{\odot}$ (see also Martins et al. \cite{martins}). Given these values, the 2.20d photometric period for Wd1-6a implies an orbital separation of $\sim28R_{\odot}$; two such supergiants could barely be accommodated in such a system, although the situation would be improved if the secondary were a less luminous $\sim$O8 V or $\sim$O9 III star ($\sim8.5R_{\odot}$ to $\sim13.7R_{\odot}$; Martins et al. \cite{martins}), as suggested by its SB1 classification (Sect. 3.1). 
One might also suppose that the true orbital period is twice this value; such a  conclusion appears inescapable for Wd1-53a, where a 1.30d period implies a separation of only $\sim20R_{\odot}$ and the double-lined nature of the spectra requires comparable bolometric luminosities for both components. The eclipsing nature of Wd1-36 implied by its lightcurve morphology requires an orbital separation of $\sim36R_{\odot}$, implying that the stars are essentially in contact.
As a consequence one might reasonably conclude that the stellar winds in such compact binary configurations  cannot accelerate to their terminal velocities and hence WCZs do not readily form. Moreover even if one assumed a smaller, less evolved companion  for e.g. the  single-lined binary Wd1-6a, such a star would support a weaker wind which might be expected to lead to reduced
 X-ray emission from the WCZ (cf. Sect. 3.1).

\subsection{Conclusions and future perspectives}

Wd1 appears to be an ideal laboratory for the study of X-ray emission from massive single and binary stars. Although significant 
interstellar reddening is present, it is not so high that soft emission is undetectable and its proximity means that individual stellar sources may be spatially  resolved. This can be contrasted with the Arches and Quintuplet, which are located within the Galactic centre region and  where only a handful of hard X-ray sources may be discerned (Wang et al. \cite{wang}). 
Moreover, at 5Myr old, Wd1 hosts a rich co-eval population of massive evolved stars that straddles the wind bistability limit, allowing the dependence of X-ray emission on wind parameters to be probed (cf. Owocki \& Cohen \cite{owocki}).

In order to exploit this potentiality, further multi-epoch optical and near-IR spectroscopy will be required to fully constrain the binary population and determine stellar and wind properties via model-atmosphere analysis (cf. Clark et al. \cite{clark18}). Deeper multi-epoch X-ray observations will also be needed in order to (i) reach currently  undetected  stellar cohorts, in particular  the O giants (to further test the $L_{\rm X}/L_{\rm bol}$ relation) and the B1-4 supergiants (to determine the nature of X-ray emission below the wind bistability limit) and (ii) search for orbital modulation in known binaries. Finally, these data will need to serve as input into individually tailored hydrodynamical simulations of the binary cohort in order to quantify the role of the WCZ and physics governing it.

Such an analysis will  lead to a better understanding of the stellar winds which mediate evolution for massive stars and the  conditions under which a WCZ forms and dominates the high energy emission of massive binaries. At this epoch at least 22 of the $>166$ massive evolved stars within  Wd1 appear strong candidates to host detectable WCZs (comprising the eight O stars with hardness colours $>-0.5$ at $>1\sigma$, the (post-) interacting binaries Wd1-9, 13 and 30a and all WR detections with the exception of  WR E; Sect. 3). These systems are of particular relevance for understanding the role colliding wind binaries play in the production of cosmic rays and the significance of this process relative to other mechanisms, such as  shocks in winds of single stars, generation in SNe remnants and pulsar winds and the interaction between cluster winds and SNe (Bednarek et al. \cite{bed}, Aharonian et al. \cite{aharonian} and refs. therein).

\begin{acknowledgements}
This research was  supported by the Science and
Technology Facilities Council and the Spanish Ministerio de Ciencia,
Innivaci\'{o}n y Universidades under grant AYA2015-68012-C2-2-P
(MINECO/FEDER).
 This research has made use of the SIMBAD database, operated
at CDS, Strasbourg, France.

\end{acknowledgements}

\appendix

\section{Appendix A - summary of the X-ray properties of massive cluster members}

\begin{table*}
\begin{center}
\caption{Summary of spectrally classified cluster members associated with X-ray emission (adapted from Clark et al. \cite{clark08}). Columns 1 and 2 present optical IDs and spectral classifications derived from Clark et al. (\cite{clark05}, \cite{clark13}, \cite{clark18} and in prep.), Crowther et al. (\cite{crowther06a}), Negueruela et al. 
(\cite{iggy10}) and Ritchie et al. (\cite{ritchie09}, \cite{ritchie10a}). X-ray designations are given in column  3 and corresponding physical data from Clark et al. (\cite{clark08}) in columns 4-6. Additional information is presented in column 7, including {\bf E}clipsing or  {\bf P}eriodic photometric modulation (Bonanos \cite{bonanos}), the  presence or absence of periodicity in the {\bf R}adial {\bf V}elocity data derived from the 2008-9 observing seasons (Ritchie et al. \cite{ritchie09}, \cite{ritchie10b}, Clark et al. in prep.), apparent SB2 binaries (Clark et al. in prep.) and non-thermal radio emission (Dougherty et al. \cite{dougherty}).  W1041 has been observed as part of our RV variability survey but the breadth of the Paschen series lines prevents analysis of RV variability  (Fig. 1; Clark et al. in prep.).}
\begin{tabular}{lcccccl}
\hline
Opt. ID   & Spectral type   & X-ray identifier    &  Counts & Hardness & Flux                           & Notes \\
          &                 &                    &         & ratio    & ($10^{-7}$ ph cm$^{-2}$ s$^{-1}$)& \\ 
\hline 
W1        & O9.5 Iab        & 164659.3-455046    & $11.3^{+ 3.6}_{- 3.6}$   & $-0.27_{-0.31}^{+0.32}$  & 4.9 &  \\
W6a       & B0.5 Iab +?     & 164703.0-455023    & $11.1^{+ 3.9}_{- 3.3}$   & $-0.47_{-0.29}^{+0.31}$ & 4.7 &  P(2.20d) \\
W9        & sgB[e]          & 164704.1-455031    &  $462.2^{+22.7}_{-22.4}$  & $0.23_{-0.05}^{+0.05}$ & 241.2 &Interacting binary \\
W10       & B0.5Ia+OB       & 164703.3-455034    & $10.2^{+ 3.8}_{- 3.5}$   & $-0.43_{-0.32}^{+0.34}$  & 4.2 & SB2, RV variable \\
W13       &B0.5 Ia$^+$/WN10h + OB Ia&164706.4-455026 & $18.9^{+ 4.1}_{- 5.2}$     & $ 0.14_{-0.24}^{+0.24}$  &13.1 & SB2, E(9.27d)   \\
W15       & O9.5 Ib         & 164706.6-455029    & $10.2^{+ 3.7}_{- 3.3}$   & $-0.62_{-0.28}^{+0.30}$  & 4.5 & Not RV variable \\
W17       & O9 Iab          & 164706.2-455048    & $7.2^{+ 4.0}_{- 3.8}$    & $-1.00^{+0.63}$          & 2.4 & Non-thermal radio, not RV variable       \\
W24       & O9 Iab +O?      & 164702.1-455112    & $19.3^{+ 5.4}_{- 5.2}$   & $ 0.14_{-0.27}^{+0.27}$  & 11.9 & SB2? RV variable \\
W25       & O9 Iab          & 164705.7-455033    & $11.2^{+ 4.0}_{- 3.6}$   & $-1.00^{+0.36}$          & 4.9 &      \\
W27       & O7-8 Ia$^+$     & 164705.1-455041    & $32.9^{+ 5.8}_{- 6.9}$   & $-0.72_{-0.14}^{+0.15}$  & 13.4 & Not RV variable, Overluminous \\ 
          &                 &                    &                          &                          &    & merger product?  \\
W30a      & O4-5 Ia$^+$     & 164704.1-455039    & $552.2^{+24.9}_{-24.5}$  & $-0.15_{-0.04}^{+0.04}$  &  253.4 & RV variable, Overluminous  binary    \\
          &                 &                    &                          &                          &        & interaction product? \\
W36       & OB Ia +  OB Ia  & 164705.0-455055    & $43.4^{+ 7.2}_{- 7.4}$   & $-0.79_{-0.11}^{+0.13}$  & 19.0 & SB2,  E(3.18d)    \\  
W38       & O9 Iab          & 164702.8-455046    & $9.0^{+ 4.2}_{- 3.6}$    & $-0.39_{-0.43}^{+0.41}$  & 3.3 & Not RV variable \\
W41       & O9 Iab          & 164702.7-455057    & $14.1^{+ 4.6}_{- 4.0}$   & $-0.21_{-0.31}^{+0.30}$  & 6.1  &      \\
W47       & O9.5 Iab        & 164702.5-455117    & $18.0^{+ 4.8}_{- 6.0}$   & $0.22_{-0.29}^{+0.30}$   & 10.1 &       \\
W53a      & OB Ia + OB Ia   & 164700.3-455131    & $56.7^{+ 7.6}_{- 8.2}$   & $-0.58_{-0.11}^{+0.12}$  & 24.0 & SB2, P(1.30d)  \\  
W56b      &  O9.5 Ib        & 164658.8-455145    & $14.5^{+ 3.9}_{- 4.2}$   & $-0.02_{-0.28}^{+0.28}$  & 6.8 &  Not RV variable      \\ 
W61b      & O9.5Iab         & 164702.5-455142    & $3.4^{+2.7}_{-3.0}$      & $-1.00^{+0.8}$           & 1.5 &       \\
W62a      & B0.5 Ib         & 164702.5-455137    &  $15.3^{+ 4.5}_{- 4.3}$   & $-0.51_{-0.26}^{+0.27}$ & 6.5 &       \\
W65       &  O9 Ib          & 164703.8-455146    &  $12.6^{+ 3.5}_{- 4.0}$   & $ 0.47_{-0.28}^{+0.26}$ & 6.8 & Low level RV variable \\ 
W74       & O9.5 Iab        & 164707.0-455012    & $9.3^{+ 3.4}_{- 3.2}$    & $-0.75_{-0.24}^{+0.29}$  & 4.0 & Not RV variable \\
W84       & O9.5 Ib         & 164659.0-455028    & $10.4^{+ 3.4}_{- 3.5}$   & $-1.00^{+0.30}$          & 3.9 & Not RV variable  \\
W232      & O9.5-B0 Iab     & 164701.4-455235    &  $11.9^{+ 3.1}_{- 4.2}$   & $-0.25_{-0.29}^{+0.30}$  & 5.4 & RV variable, C07-X6 \\
W1005     & B0 Iab          & 164654.2-455154    & $9.7^{+ 3.0}_{- 3.6}$    & $-0.54_{-0.29}^{+0.32}$  & 6.4 & Not RV variable, C07-X7       \\
W1018          & O9.5 Iab        & 164658.2-455056    &  $12.7^{+ 3.5}_{- 4.1}$   & $-0.18_{-0.29}^{+0.30}$  & 6.5 &      \\
W1027     &  O9.5 Iab       & 164701.0-455006    &  $14.5^{+ 3.9}_{- 4.1}$   & $-0.43_{-0.25}^{+0.26}$  & 6.2 &      \\
W1033      & O9-9.5 I-III    & 164702.3-455233    & $9.9^{+ 2.8}_{- 3.8}$    & $-0.54_{-0.28}^{+0.31}$  & 4.9  & Not RV variable, C07-X5      \\
W1036    &  O9.5 Ib        & 164702.7-455212    & $12.3^{+ 3.8}_{- 3.7}$   & $-0.02_{-0.31}^{+0.30}$  & 5.6 & Not RV variable, C07-X4 \\        
W1040      &  O9.5 Iab/b     & 164704.5-455008    & $14.2^{+ 4.2}_{- 3.9}$   & $-0.42_{-0.25}^{+0.27}$  & 6.3 & RV variable, C07-X3 \\
W1041     &  O9.5 Iab +?    & 164704.4-455109    & $13.9^{+ 3.8}_{- 4.8}$   & $ 0.02_{-0.31}^{+0.31}$ & 6.6 & Broad Paschen lines, SB2? \\
W1051      & O9 III          & 164706.9-454940    & $12.6^{+ 3.6}_{- 4.0}$   & $-0.33_{-0.28}^{+0.29}$  & 5.4 & Not RV variable         \\
W1055      & B0 Ib           & 164707.8-455147    & $14.5^{+ 3.9}_{- 4.2}$    & $0.40_{-0.27}^{+0.25}$  & 7.2 & Not RV variable  \\
W1064      & O9.5 Iab        & 164711.5-455000    & $5.1^{+ 3.1}_{- 2.5}$    & $ 1.00_{-0.49}$          & 2.3 &      \\
WR A      &  WN7b           & 164708.3-455045    & $743.7^{+27.3}_{-27.3}$  & $ 0.19_{-0.04}^{+0.04}$  & 380.7 & P(7.63d)     \\
WR B      & WN7o            & 164705.3-455104    & $237.5^{+16.1}_{-16.5}$  & $0.21_{-0.07}^{+0.06}$   & 133.2 & E(3.51d) \\
WR D      & WN7o            & 164706.2-455126    & $14.3^{+ 4.7}_{- 4.6}$   & $ 1.00_{-0.19}$          & 16.6  &       \\
WR E      & WC9             & 164705.9-455208    & $8.0^{+ 4.0}_{- 3.3}$    & $-0.57_{-0.40}^{+0.39}$  & 3.7  & RV binary?      \\
WR F      & WC9d            & 164705.2-455224    & $109.6^{+10.9}_{-11.3}$  & $ 0.63_{-0.08}^{+0.08}$  & 65.7 & RV(5.05d), possible trinary? \\
WR G      & WN7o            & 164704.0-455124    & $19.2^{+ 5.5}_{- 5.1}$   & $0.55_{-0.25}^{+0.25}$   & 17.5 &       \\
WR L      & WN9h:           & 164704.1-455107    & $80.8^{+ 9.6}_{-10.3} $  & $0.34_{-0.11}^{+0.11}$   & 48.8 & RV($\sim54$d) \\
WR N      & WC9d            & 164659.8-455525    & $11.3^{+ 3.7}_{- 3.6}$   & $ 0.57_{-0.29}^{+0.25}$  & 19.7 & Dust producing CWB?      \\
WR O      & WN6o            & 164707.6-455235    &  $54.5^{+ 7.7}_{- 8.1}$   & $ 0.07_{-0.14}^{+0.14}$  & 27.0 &      \\
WR R      & WN5o            & 164706.0-455022    & $30.8^{+ 5.9}_{- 6.8}$   & $ 0.27_{-0.20}^{+0.20}$  & 20.5  &       \\
WR U      & WN6o            & 164706.5-455039    &   $161.2^{+13.6}_{-13.2}$  & $-0.14_{-0.08}^{+0.08}$  & 81.4 &     \\
WR W      & WN6h            & 164707.6-454922    &   $21.1^{+ 5.2}_{- 4.8}$   & $ 0.72_{-0.19}^{+0.16}$  & 15.9  &     \\
\hline
\end{tabular}
\end{center}
\end{table*}

\end{document}